\documentclass[eqsecnum,twocolumn,tightenlines,floats,floatfix,prc,nofootinbib]{revtex4-1}

\def\st#1{{\kern-4pt} \not\!#1}

\def\bea{\begin{eqnarray}}
\def\eea{\end{eqnarray}}

\def\be{\begin{equation}}
\def\ee{\end{equation}}
\def\ba{\begin{eqnarray}}
\def\ea{\end{eqnarray}}

\usepackage{graphics}
\usepackage{graphicx}
\usepackage{epsf} 
\usepackage{amsmath}
\usepackage{amssymb}
\usepackage{slashed}
\usepackage{bbold}
\usepackage{bm}
\usepackage{color}

\def\sfrac#1#2{{\textstyle \frac{#1}{#2}}}

\begin{document}

\phantom{0}
\hspace{5.5in}\parbox{1.5in}{ \leftline{JLAB-THY-12-1481}
                \leftline{}\leftline{}\leftline{}\leftline{}
}

\title{\bf 
Covariant nucleon wave function with S, D, and P-state components \\ \phantom{0}}



\author{
Franz Gross$^{1,2}$, G.~Ramalho$^{3}$ and 
M.~T.~Pe\~na$^{3}$ }
\vspace{-0.1in}

\affiliation{
$^1$Thomas Jefferson National Accelerator Facility, Newport News, 
Virginia 23606, USA \vspace{-0.15in}}
\affiliation{
$^2$College of William and Mary, Williamsburg, Virginia 23185, USA
\vspace{-0.15in}}
\affiliation{
$^3$Universidade T\'ecnica de Lisboa,
CFTP, Instituto Superior T\'ecnico,
Av.\ Rovisco Pais, 1049-001 Lisboa, Portugal }

\phantom{0}

\begin{abstract}
Expressions 
for the nucleon wave functions in the 
covariant spectator theory (CST) are derived. 
The nucleon is described 
as a system with a off-mass-shell 
constituent quark, free to interact with an external probe, and 
two spectator constituent quarks on their mass shell. 
Integrating over the internal momentum 
of the on-mass-shell quark pair allows us to derive an effective  nucleon wave function that can be written only in terms 
of the quark and diquark (quark-pair) 
variables.
The derived nucleon wave function includes contributions from  S, P and D-waves.
\end{abstract}

\vskip 1cm
\date{\today}
\maketitle

\vspace{-1.5cm}


\section {Introduction}

In this work we use the covariant spectator theory (CST) \cite{Gross:1969rv,Nucleon,Octet,Omega,ExclusiveR}  to determine the structure of the valence quark contributions to the wave functions of the nucleon.  In the CST, baryon systems consist of an off mass-shell 
constituent quark, free to interact with electromagnetic fields or other probes, 
and two noninteracting on mass shell constituent quarks that are spectators to the interaction.   Since the interaction does not depend on the internal momentum of these on-shell spectators, we can integrate over their internal momentum and express the effective matrix element in terms of a nucleon composed of an off-shell quark and an on-shell  quark pair (or diquark) with an average mass $m_s$, which becomes a parameter in the wave function.  

Previously we have assumed a pure S-wave structure for the wave functions; here we add contributions associated with D and P-wave components.  We begin by discussing the general form of the CST matrix elements.  Then, for each angular momentum component, nonrelativistic  wave functions are constructed first, and then generalized to relativistic form.

\section{CST matrix elements} \label{sec:II}

\subsection{Relativistic impulse approximation} \label{sec:IIA}

\begin{figure}
\centerline{
\mbox{
\includegraphics[width=3.5in]{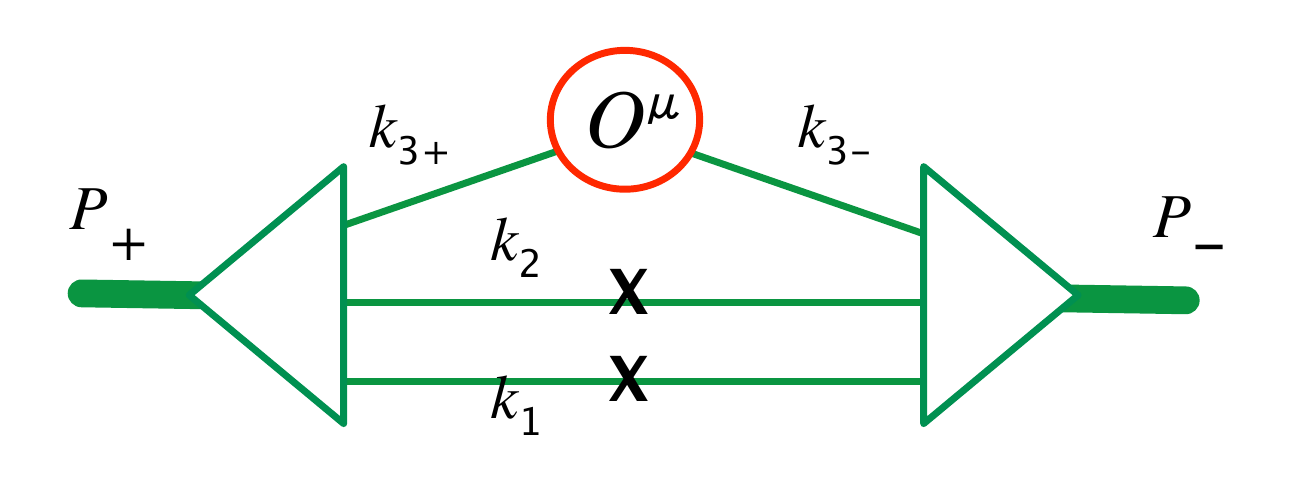} }}
\caption{\footnotesize{(Color on line) Diagrammatic representation of the RIA approximation to the CST matrix element of the operator $O^\mu$.  The quarks with four moments $k_1$ and $k_2$ are on-shell spectators (represented by the $\times$). }}
\label{fig:1}
\end{figure}

In the relativistric impulse approximation (RIA), it is assumed that the interaction is well described by a single quark operator $O^\alpha$, and that interactions involving a pair of quarks  
(i.e.~exchange or interaction currents) can be neglected.  In this case, the baryonic matrix elements in the CST can be written (for a brief discussion of corrections to the RIA, see Ref.~\cite{VanOrden:1995eg})
\bea
{\cal O}^\mu_{\lambda_+\lambda_-}(P_+,P_-)
=\sum^3_{i=1}\sum_{\lambda_j\lambda_\ell}\int_{k_jk_\ell} \left< O^\mu\right>_i \label{eq:2.1a}
\eea
where the first sum is over all three possible choices for the interacting quark, and the three-momentum integrations and helicity sums are over the momenta and helicities of the on-shell spectator quarks, with $i,j,\ell$ in cyclic order, so that, for example, if the third quark is interacting, then the (12) pair are the spectators.  The matrix element is
\bea
\left< O^\mu\right>_i=\overline{\Psi}_{\lambda_j\lambda_\ell,\lambda_+}(P_+,k_jk_\ell) 
&&O^\mu \Psi_{\lambda_j\lambda_\ell;\lambda_-}(P_-,k_jk_\ell)\qquad
\eea
with $P_-$ ($P_+$) and $\lambda_-$ ($\lambda_+$) the four-momenta and helicity of the incoming (outgoing) baryon, and $\Psi$ the baryon wave function.  We have suppressed the Dirac indices of the off-shell quark.  The matrix element is illustrated in Fig.~\ref{fig:1}.   In the CST the spectator quarks are constrained to their positive energy mass-shell, so the covariant volume integral is (with $j=1$ and $\ell=2$)
\bea
\int_{k_1k_2}&&\equiv \int\frac{d^4k_1d^4k_2}{(2\pi)^6}\delta_+(m_1^2-k_1^2)\delta_+(m_2^2-k^2_2)
\nonumber\\&&
=\int \frac{d^3k_1\,d^3k_2}{(2\pi)^6 4E_1E_2} \label{eq1.29a}
\eea
where the four-momenta of the on-shell quarks are
\bea
k_1&=&\{E_{1}, {\bf k}_1\}
\nonumber\\
k_2&=&\{E_{2}, {\bf k}_2\}\, ,
\eea
with $E_i=\sqrt{m^2_i+{\bf k}_i^2}$.  

In this paper we will first assume the dressed masses of the three quarks are equal, so that $m_1=m_2=m_3=m$.   (Later we will consider the case when $m_u\ne m_d$.) 
Since the color factor (suppressed) is fully antisymmetric, when the particles are identical   this means that the remaining matrix element must be symmetric under the interchange of the three quarks.  In this case the three terms for different $i$ are identical, and we may write the full result as three times the result for the (arbitrary) choice of $k_1$ and $k_2$ as spectators transforming (\ref{eq:2.1a}) to
\bea
{\cal O}^\mu_{\lambda_+\lambda_-}(P_+,P_-)
=3\sum_{\lambda_1\lambda_2}\int_{k_1k_2} \left< O^\mu\right>_3 .\label{eq:2.1}
\eea
This means that it is only necessary for the wave function $\Psi(P,k_1,k_2)$ to be symmetric under the interchange of quarks 1 and 2; symmetry under  the interchanges of the other quarks, $1\leftrightarrow 3$ or $2\leftrightarrow 3$, was used when the full result (\ref{eq:2.1a}) is simplified to (\ref{eq:2.1}).

\subsection{Relativistic definition of the quark-diquark wave function} 
\label{sec:IIB}

As shown in Fig.~\ref{fig:1}, the dependence of the matrix element on the relative momentum of the two on-shell quarks is determined by the wave function only, and not by the structure of the operator $O^\mu$, which depends only on the momenta $k_{3\pm}$ of the initial and final off-shell quark
\bea
k_{3\pm}&&=P_\pm-(k_1+k_2).
\eea
Since the wave function will be determined phenomenologically, very little is lost by averaging over 
the relative momenta of the two on-shell quarks and introducing a new wave function that depends only on the {\it total\/} four-momentum of the on-shell quark pair (which will be called a  diquark).   In order to do this we introduce the diquark momentum variables
\bea
k&=&(k_1+k_2)\nonumber\\
r&=&\sfrac12(k_1-k_2), 
\label{eqRK}
\eea
which, together with the fixed total momentum $P=k_1+k_2+k_3$, are a complete set. (If $k\to k_1+k_2-\frac23 P$, these would be the Jacobi coordinates.)  
The  diquark four-momentum $k=k_1+k_2$ has a mass $s=(k_1+k_2)^2$,  which is initially unconstrained.  In this notation, the  integrals (\ref{eq1.29a}) can be re-expressed  as integrals over ${\bf k}$, $s$, and the direction of the internal 
relative three momentum ${\bf r}=\sfrac12({\bf k}_1-{\bf k}_2)$ 
of the diquark 
\bea
\int_{s\,k}&=&\underbrace{\int \frac{d\Omega_{\hat{\bf r}}}{4(2\pi)^3}\int_{4m_q^2}^\infty ds\sqrt{\frac{s-4m_q^2}{s}} }_{\int_{s}}\underbrace{\int\frac{d^3k}{(2\pi)^32E_s}}_{\int_k},\qquad\quad \label{eq1.30}
\eea
where $m_q$ is the dressed quark mass, and 
$E_s=\sqrt{s+{\bf k}^2}$ is the energy of a diquark of mass $\sqrt{s}$, 
and the angular integrals $d\Omega_{{\bf r}}$ 
are written in the {\it rest frame of the diquark\/}.  
The relation (\ref{eq1.30}) is discussed in Appendix \ref{apDiquark}.

Using the relation (\ref{eq1.30}) the matrix element (\ref{eq:2.1}) can be written
\bea
{\cal O}^\mu_{\lambda_+\lambda_-}&&(P_+,P_-)
\nonumber\\&&
=3\sum_{\Lambda}\int_{k} \overline{\Psi}_{\Lambda\lambda_+}(P_+,k)O^\mu\Psi_{\Lambda\lambda_-}(P_-,k)\qquad \label{eq:2.8}
\eea
where the sum over $\Lambda$ includes all possible polarization pairs $\lambda_1, \lambda_2$ of the diquark
, and the new density in (\ref{eq:2.8}) is related to the density in (\ref{eq:2.1})   by
\bea
\sum_{\lambda_1\lambda_2}\int_{s}&&\overline{\Psi}_{\lambda_1\lambda_2,\lambda_+}(P_+,k_1k_2) 
\otimes \Psi_{\lambda_1\lambda_2;\lambda_-}(P_-,k_1k_2)
\nonumber\\
&&\equiv\sum_{\Lambda}\; \overline{\Psi}_{\Lambda\lambda_+}(P_+,k)\otimes \Psi_{\Lambda\lambda_-}(P_-,k)\Big|_{s=m_s^2}\, , \qquad\label{eq:2.9}
\eea 
where the operator $\otimes$ shows where the Dirac operator $O^\mu$ is to be inserted.  This equation gives the precise relationship between the quark-diquark wave function, denoted by  $\Psi_{\Lambda,\lambda}(P,k)$, and the full three-quark wave function, $\Psi_{\lambda_1\lambda_2,\lambda}(P,k_1k_2)$.  (Note that these two wave functions are distinguished from each other  only by their list of arguments.)    Equation (\ref{eq:2.9}) shows that the quark-diquark wave function is obtained from the full three-quark wave function by averaging over the directions of the relative three momentum ${\bf r}$ (in the rest system) and replacing the integral over the continuous diquark mass, $\sqrt{s}$,  by its mean value, $m_s$, which now becomes a parameter of the theory.  All remaining factors from the integral $\int_s$ are absorbed into the normalization of the wave function.  After the average has been carried out, the diquark energy becomes $E_s= \sqrt{m_s^2+ {\bf k}^2}$.

We emphasize that, in the RIA, the replacement (\ref{eq:2.9}) is exact, as long as the effective diquark mass $m_s$ is treated as a parameter.  Later we will constrain the values of this parameter, and in so doing we make an approximation.

\subsection{Nonrelativistic definition of the quark-diquark wave function}

The nonrelativistic limit of (\ref{eq1.30}) is obtained by assuming the quark and diquark masses are very large, and absorbing the mass factors into the wave function normalization.  
In this case
\bea
s&&=\left(\sqrt{m_q^2+k_1^2}+\sqrt{m_q^2+k_2^2}\right)^2-{\bf k}^2
\nonumber\\
&&\simeq 4m_q^2+ 4 {\bf r}^2.
\label{eqSNR}
\eea
For very large masses 
($E_s \to m_s$; $s \to 4 m_q^2$),
we can write 
\bea
m_q m_s\int_{s\,k}&&\to \frac1{16}
\int\frac{d^3k}{(2\pi)^3}\int \frac{d\Omega_{\hat{\bf r}}}{(2\pi)^3}\int_{4m_q^2}^\infty ds\sqrt{s-4m_q^2}
\nonumber\\
&&=\int \frac{d^3k}{(2\pi)^3}\int\frac{d^3r}{(2\pi)^3},
\label{eqINTNR}
\eea
where, in the last step, the integral over $s$ has been replaced by an integral over  the magnitude of the relative momentum, $r=|{\bf r}|$, using 
(\ref{eqSNR}).

The result (\ref{eqINTNR}) 
also follows directly from the transformation from ${\bf k}_1, {\bf k}_2$ to ${\bf k}, {\bf r}$, which gives   
\bea
\int\int\frac{ d^3k_1\,d^3k_2}{(2\pi)^6}=\int \frac{d^3k}{(2\pi)^3}\int\frac{d^3r}{(2\pi)^3}\equiv \int_{k,r}^{\rm NR}. \qquad\label{eq:2.12}
\eea
Hence the nonrelativistic definition of the diquark wave function is identical to (\ref{eq:2.9}), with
\bea
  \frac{m_q}{2} \int_s\to \int\frac{d^3r}{(2\pi)^3}\, .
\eea

Since the details of how the averaging over the internal diquark structure depends on the individual angular momentum components of the wave function, this discussion is postponed until Sec.~\ref{sec:IV}.

\section{Electromagnetic interaction and normalization of the wave function} \label{sec:III}

The interaction of the photon with a constituent quark 
is decomposed into Dirac and Pauli components.   
\be
j^\mu (Q^2)=  j_1 \left(\gamma^\mu -\frac{\st q q^\mu}{q^2}
\right) + j_2 \frac{i \sigma^{\mu \nu} q_\nu}{2M}, \label{eq:3.1}
\ee
where $M$ is the nucleon mass.  Note that the constituent quarks 
can have anomalous magnetic moments, and that we add the term $-\slashed{q}q^\mu/q^2$ to insure that the current is always conserved \cite{Batiz:1998wf}.  In this paper we will limit discussion to $u$ ($I_z\equiv t=+\frac12$) and $d$ ($t=-\frac12$) quarks only, in which case the quark current $j^\mu$ and the quark form factors $j_i$ are operators in isospin space.  The Dirac  ($j_1$) and Pauli ($j_2$) form factors  are therefore decomposed into two 
independent isospin structures
\be
j_i(Q^2)= \sfrac{1}{6} f_{i+}(Q^2) + \sfrac{1}{2} f_{i-}(Q^2) \tau_3,
\ee
where $\tau_3$ is the $z$-projection of the isospin operator.
This equation   
defines the isoscalar and isovector form factors 
that can also be expressed as the $u$ and $d$ 
quark form factors.
In the CST quark model the 
functions $f_{i \pm}(Q^2)$ are parametrized 
using a vector meson dominance representation \cite{Nucleon}.



Using (\ref{eq:2.8}) and the operator (\ref{eq:3.1}), the baryon form factors become
\bea
J_{\lambda_+ \lambda_-}^\mu&&(P_+,P_-)
\nonumber\\&&
= 3\sum_\Lambda\int_{k} \overline{\Psi}_{\Lambda\lambda_+}(P_+,k) 
j^\mu(Q^2) \Psi_{\Lambda\lambda}(P_-,k),\qquad
\label{eqCurrent1}
\eea
%
%
For the nucleon, at 
$Q^2=0$, when $P_+=P_-= (M,0,0,0)$ and 
\bea
j^0(0)= e_0 \Big(\sfrac16+\sfrac12\tau_3\Big)\gamma^0=j_1(0)\gamma^0\,  \label{eq:3.4}
\eea
(because of the presence of the factor $e_0$, this operator at $Q^2=0$ gives the renormalized quark charge),  we require that the wave function be diagonal in the nucleon polarization and normalized to the correct nucleon charge:  
\be
J^0_{\lambda_+\lambda_-}(P,P)\equiv Q= \sfrac{1}{2} (1 + \tau_3)\,\delta_{_{\lambda_+,\lambda_-}}.\label{eqCharge}
\ee 
It is always possible to normalize the wave functions so that the proton charge is unity, but to obtain zero for the neutron charge places constraints on the structure of the wave function.

\section{Wave functions of the nucleon} \label{sec:IV}

\subsection{Total relativistic wave function}

The total relativistic quark-diquark wave function will 
be written  as a sum of S, P, and D-wave components
\bea
&&\Psi_{\Lambda\lambda}(P,k ) 
\nonumber\\&&
=n_S \Psi^{S}_{\Lambda\lambda}(P,k)+
n_P \Psi^{P}_{\Lambda\lambda}(P,k)+n_D\Psi^{D}_{\Lambda\lambda}(P,k )  \qquad\label{psiRel} 
\eea
where $\Psi^{L}_{\Lambda\lambda}(P,k)$ is the $L$=S, P, or D-state 
wave function 
to be defined in the next subsections.

We will require that {\it each\/} of these components be separately normalized to the nucleon charge, Eq.~(\ref{eqCharge}).   This requirement will be decisive in fixing the structure of the wave functions. When each component has been separately normalized, the  S-state normalization constant, $n_S$, 
will be determined by the P-state mixing parameter, $n_P$ and the
D-state mixing parameter $n_D$
\bea
n_S=\sqrt{1-n_P^2-n_D^2}\, .
\eea

\subsection{S-state component} \label{sec:Sstate}

\subsubsection{Spin-isospin part of the nonrelativistic wave function}

The construction of the nonrelativistic S-state proton 
wave function was given previously \cite{Nucleon}, but here the construction will done differently.  We obtain the same result as we did before, but this way of doing things will be helpful 
when we extend the discussion to the D-state.

The wave function is written as a product of a momentum space 
wave function symmetric in $k_1$ and $k_2$ (this is the only possibility for an S-state) and a spin-flavor part also symmetric under the interchange of quarks 1 and 2.   There are two possible structures that contribute to the symmetric spin-flavor part: the (0,0) component which is a direct product of an operator antisymmetric in the flavor (isospin 0) and antisymmetric in the spin (spin 0), and the (1,1) component symmetric in the flavor (isospin 1) and symmetric in the spin (spin 1).  The isospin 0 and 1 components can be written in the following form 
\ba
{\bm \phi}^0 \chi^t&&= \chi^{t}
\nonumber\\
{\bm \phi}_{\ell}^1\, \chi^t&&= 
- \frac{1}{\sqrt{3}}
\left( \tau \cdot \xi^{\ast}_\ell \right) \chi^{t}
\label{eqFlavor}
\ea
with $\chi^t$ the two-component isospinor of the nucleon
\bea \chi^{+\frac{1}{2}}= \left(\begin{array}{c} 1 \cr 0 \end{array} \right)\qquad 
\chi^{-\frac{1}{2}}= \left(\begin{array}{c} 0 \cr 1 \end{array} \right),
\eea
and  $\xi_\ell$ the isospin vector of the isospin-one diquark, defined in the usual way (in rectangular coordinates)
\bea
\xi_\pm= \mp\frac1{\sqrt{2}}
\left(
\begin{array}{r}
  1   \\
  \pm i   \\
  0   
\end{array}
\right),\qquad \xi_0= 
\left(
\begin{array}{r}
  0   \\
  0   \\
  1   
\end{array}
\right)\, . \label{eq:4.10}
\eea
(This notation differs slightly from Ref.~\cite{Nucleon}, but the results are the same.)  The spin one vectors are normalized to
\bea
(\xi_{\ell'})^\dagger\cdot \xi_\ell&&=\delta_{\ell',\ell}
\nonumber\\
\sum_\ell(\xi_{\ell})_i(\xi_\ell)^\dagger_j&&=\delta_{ij}\, 
\eea
where the $i$th component of the vector $\xi_\ell$ is denoted $(\xi_\ell)_i$.  In the nonrelativistic limit, the spin wave functions and operators have precisely the same form.

Using the operator notation (\ref{eqFlavor}) the total wave function of the nucleon can now be written in a general form which displays its dependence on all three quark momenta, but still describes the flavor and spin of the three quarks in terms of the quark-diquark description already developed. 
By separating explicitly the spin 0 isoscalar diquark contribution, 
$\Psi_\lambda^{S,0}$, from the spin 1 vector diquark contribution, 
$\Psi^{S,1}_{\Lambda\lambda}$, we have
\bea
\Psi^S_{\Lambda\lambda}({\bf P};{\bf k}_1,{\bf k}_2)
=&&\,\cos\theta_S\,{\bm\phi}^0\,\Psi_\lambda^{S,0}({\bf P};{\bf k}_1,{\bf k}_2) 
\nonumber\\
&&+ \sin\theta_S\,{\bm \phi}^1\, \Psi^{S,1}_{\Lambda\lambda}({\bf P};{\bf k}_1,{\bf k}_2)\qquad\quad
\label{eqPsiSymm2}
\eea
where $\lambda$ is the spin projection of the nucleon, 
$\Lambda$ the spin projection of the spin-1 diquark, 
and $\Psi^{S,s}_{\Lambda\lambda}({\bf P};{\bf k}_1,{\bf k}_2)$ 
is the S-state spin-space part 
of the wave functions (with $s$ the spin of the diquark and no $\Lambda$ dependence in the spin-0 component).  
Explicitly
\bea
\Psi_\lambda^{S,0}({\bf P};{\bf k}_1,{\bf k}_2)=&& \Phi_S({\bf P};{\bf k}_1,{\bf k}_2)\left|\sfrac12,\lambda\right>
\nonumber\\
\Psi_{\Lambda\lambda}^{S,1}({\bf P};{\bf k}_1,{\bf k}_2) 
=-\sfrac1{\sqrt{3}} &&\Phi_S({\bf P};{\bf k}_1,{\bf k}_2)
\sigma\cdot\varepsilon_\Lambda^*\left|\sfrac12,\lambda\right>\,  
,\qquad\label{eq:4.15}
\eea
where we have introduced our notation for the diquark spin functions, modeled after the isospin operators (\ref{eqFlavor}), with $\varepsilon_\Lambda$ the spin vector for the spin-1 diquark [with its components in spin-space defined as in (\ref{eq:4.10})], and $\left|\sfrac12,\lambda\right>$ the initial nucleon spin state with polarization $\lambda$.  At this point we assume that the same momentum space wave function, $\Phi_S$, accompanies both of the spin parts of the wave function; later this assumption will be relaxed as the wave function is determined from phenomenological fits to the DIS data. We emphasize that, at this stage, the two components of the wave function have spins equal to their isospins, but it is best to label these components by their spin because later (in Sec.~\ref{udmix}) we will break the isospin invariance.

Using this wave function, the nonrelativistic matrix element of the charge operator, (\ref{eqCharge}) becomes
\bea
Q=&&\sfrac12(1+\tau_3)\,\delta_{_{\lambda_+,\lambda_-}}
\nonumber\\
=&&\Big\{\cos^2\theta_S\,j^A+\sin^2\theta_S\,j^S\Big\} \delta_{\lambda_+,\lambda_-}
e_0
{\cal N}_S \label{eq:4.14}
\eea  
where the isospin matrix elements are
\bea
j^A &&\equiv 3
\left({\bm\phi}^0 \right)^\dagger \!  j_1(0) \,{\bm\phi}^0= e_0\left(\sfrac12+\sfrac32\tau_3\right)\nonumber\\
j^S 
&&\equiv 3\sum_\ell
\left({\bm\phi}^1_\ell\right)^\dagger \! j_1(0) \,{\bm\phi}^1_\ell 
=
\sum_\ell (\xi_\ell\cdot\tau) \,j_1(0)  \, (\xi^*_\ell\cdot\tau)  
 \nonumber \\
&&= \sfrac12 e_0\left(1-\tau_3\right)\, , \label{eq:4.16}
\eea 
where $j_1(0)$ is the quark charge operator (\ref{eq:3.4}), the normalization integral is 
\bea
{\cal N}_S= \int\frac{d^3k}{(2\pi)^3}\int\frac{d^3r}{(2\pi)^3}\,
\Phi^2_S({\bf 0};{\bf k}_1,{\bf k}_2),  \label{eq:4.16b}
\eea
and we used the fact that the spin operators in (\ref{eq:4.15}) are renormalized.  Choosing ${\cal N}_S=1/e_0$, the charge operator (\ref{eq:4.14}) becomes
\bea
Q&&=\Big\{\cos^2\theta_S\Big[\sfrac12+\sfrac32\tau_3\Big]  +\sin^2\theta_S\Big[\sfrac12-\sfrac12\tau_3 \Big]\Big\}\delta_{_{\lambda_+,\lambda_-}}
\nonumber\\
&&=\sfrac12\Big\{1+\tau_3\Big[3\cos^2\theta_S-\sin^2\theta_S\Big] \Big\} \delta_{_{\lambda_+,\lambda_-}}\, .
\eea
Hence an equal mixture of diquark spin-0 
and spin-1 components 
($\theta_S=\pi/4$), which we have used in our previous work, 
is required by the demand that the neutron charge be zero.
We will find this requirement useful in the construction of the D-state below.

The integral (\ref{eq:4.16b}) provides an explicit example of how the quark-diquark wave function emerges from an average over the internal momenta of the diquark, as shown in general in Eq.~(\ref{eq:2.9}).   Since the momentum wave function must be symmetric in $k_1$ and $k_2$, we may choose its argument to be
\bea
\chi_{nr}(k,P,r)&&=({\bf k}_1-\sfrac13{\bf P})^2+({\bf k}_2-\sfrac13{\bf P})^2
\nonumber\\
&&=
\sfrac12 {\bf k}^2+2\,{\bf r}^2-\sfrac23{\bf k}\cdot{\bf P}+\sfrac29 {\bf P}^2 
\nonumber\\
&&=
\sfrac12\left({\bf k}-\sfrac23{\bf P}\right)^2+2 {\bf r}^2 
\label{nrarg}
\eea
With this choice, the quark-diquark wave function in an arbitrary frame is defined by the relation
\bea
 \int\frac{d^3k}{(2\pi)^3}&&\int\frac{d^3r}{(2\pi)^3}\,
 \Phi_S[
\chi_{nr}(k,P_+,r)]  \Phi_S[ \chi_{nr}(k,P_-,r)]
\nonumber\\
=&&\int\frac{d^3k}{(2\pi)^3}\,  \Phi_S[\chi_{nr}(k,P_+,\bar r)]  \Phi_S[ \chi_{nr}(k,P_-,\bar r)]
\nonumber\\
\equiv&&\int\frac{d^3k}{(2\pi)^3}\, \phi_S({\bf P}_+,{\bf k})
\phi_S({\bf P}_-,{\bf k}),\qquad \label{eq:4.17}
\eea
where, in the second step we replace $r$ by and average value $\bar  r$, and the last step gives the precise relation between the quark-diquark wave function, $\phi_S({\bf P},{\bf k})$, and the three-quark wave function, $\Phi_S({\bf P};{\bf k}_1,{\bf k}_2)$, analogous to the definition given in Eq.~(\ref{eq:2.9}). 
With this definition, the normalization of the quark-diquark wave function is
\bea
e_0 \,{\cal N}_S\equiv e_0 \int\frac{d^3k}{(2\pi)^3}\, \phi_S^2({\bf 0},{\bf k})=1.
\eea
Note that the normalization of the wave function compensates for the renormalization of the quark charge, ensuring that the proton charge is correct.

\subsubsection{Relativistic wave function}

The relativistic generalization of the nonrelativistic wave function 
is straightforward, and has been discussed extensively 
in Ref.~\cite{Nucleon}.  The wave function we use is
\bea
\Psi_{\Lambda\lambda}^S(P,k)&=& \frac{1}{\sqrt{2}} 
\left[{\bm \phi}^0 u(P,\lambda) - 
{\bm \phi}^1 (\varepsilon_{\Lambda}^\alpha)^\ast U_\alpha(P,\lambda) \right] \nonumber\\
&&\qquad \times\;\psi_S(P,k).
\label{eqPsiN}
\eea
where the {\it relativistic quark-diquark\/} wave function, $\Psi^S_{\Lambda\lambda}(P,k)$, is distinguished from the {\it nonrelativistic, three-quark\/} wave function, $\Psi^S_{\Lambda\lambda}(P;k,r)$ only by its arguments, and $U_\alpha$ is
\bea
U_\alpha(P,\lambda)=
\frac{1}{\sqrt{3}} \gamma_5 \left(
\gamma_\alpha - \frac{P_\alpha}{M} \right) u(P,\lambda), \label{eq2.8}
\eea
%
with 
$u(P,\lambda)$ the free nucleon spinor with four-momentum $P$ 
and spin projection $\lambda$, $\varepsilon_{\Lambda}$ 
is the polarization vector of the spin-1 diquark state 
with polarization $\Lambda$ in the direction of ${\bf P}$ 
(and subject to the constraint $\varepsilon^\mu_\Lambda P_\mu=0$), 
and the ${\bm\phi}^{0,1}$ are the flavor wave functions of the quarks 
in a (12)3 configuration with the (12) pair 
in an isospin zero or one state [see Eq.~(\ref{eqFlavor})].  
The polarization vectors $\varepsilon^\mu$ are 
discussed in detail in Refs.~\cite{Nucleon,FixedAxis}.  The states have been normalized so the superposition (\ref{eqPsiN}) corresponds to an equal mixture of (0,0) and (1,1) components, as discussed above.

The spatial part of the S-state wave function will be fixed by comparison with the deep inelastic scattering (DIS) data as discussed in the accompaning paper \cite{following}.  The model used in earlier work had the form
\bea
\psi_S(P,k)=\frac{N_0}{m_s(\beta_1+\chi)(\beta_2+\chi)}
\eea
where
\bea
\chi=\frac{(M-m_s)^2-(P-k)^2}{Mm_s}=\frac{2P\cdot k}{Mm_s}-2\qquad
\eea
and $\beta_2>\beta_1>0$ are range parameters with the 
normalization constant $N_0$ chosen so that 
\bea
1=e_0\int_k |\psi_S(P,k)|^2\, , \label{eq:4.25}
\eea
with the covariant integration defined by 
Eq.~(\ref{eq1.30}) (with $s=m_s^2$).  Note that, in the nonrelativistic limit, $\chi$ reduces to
\bea
\chi&&=2\sqrt{1+\frac{{\bf k}^2}{m_s^2}}\sqrt{1+\frac{{\bf P}^2}{M^2}}
-2\frac{{\bf k}\cdot{\bf P}}{Mm_s}-2
\to 
\left(
\frac{{\bf k}}{m_s}-\frac{{\bf P}}{M}
\right)^2
\nonumber\\
&&\to
\frac1{4m_q^2}\left({\bf k}-\frac23{\bf P}\right)^2=
\frac1{2m_q^2}\chi_{nr}(k,P,0)  
\qquad 
\eea   
where in the last step, to display the relation between $\chi$ and $\chi_{nr}$,  we made the approximation $m_s\simeq2m_q$ and $M\simeq3m_q$, true when all momenta (and binding energies) are much smaller that $m_q$.  For future reference we note now that 
\bea
\tilde k^2&&=\left(k-\frac{(P\cdot k) P}{P^2}\right)^2=m_s^2-\frac{(P\cdot k)^2}{M^2}
\nonumber\\
&&=-m_s^2(\chi +\sfrac14\chi^2)\, .
\eea

With the normalization (\ref{eq:4.25}) the nucleon charge is
\bea
Q&=&3\sum_\Lambda\int_{k} \overline{\Psi}_{\Lambda\lambda}^S(P,k) j_1(0) \gamma^0 \Psi_{\Lambda\lambda}^S(P,k),
\eea
where $j_1(0)$ is the quark charge operator defined in Eq.~(\ref{eq:3.4}).  
This agrees with Eq.~(2.14) of Ref.~\cite{NDeltaD}  
(with the sum over the diquark  polarization $\Lambda$ extended 
to include the spin 1 diquark with polarization $\Lambda$ and the spin 0 
diquark as a separate term). 

We now discuss the construction of the D-state wave function.

\subsection{D-state component}

\subsubsection{Three-quark nonrelativistic wave function}


In order to produce a spin 1/2 state through the coupling of a D-wave operator with a state of total quark spin $S$,  the quark spin $S$ must be 3/2.
This is a purely symmetric spin state.  Hence, maintaining the requirement that the wave function be symmetric under the interchange of quarks 1 and 2 (recall that the other symmetries are handled as discussed in Sec.~\ref{sec:IIA}), the flavor-space part of the wave function can be constructed from the sum of only two components.  One will be the product of components antisymmetric under 12 interchange in both flavor and momentum space, and the other the product of symmetric components.  In analogy with Eq.~(\ref{eqPsiSymm2}), the flavor-space wave function will then be a superposition  
\bea
\Psi^D_{m_\ell}({\bf P};{\bf k}_1,{\bf k}_2)=&&\cos\theta_D\,{\bm\phi}^0 \Psi^{Da}_{m_\ell}({\bf P};{\bf k}_1,{\bf k}_2)
\nonumber\\
&&+\sin\theta_D \,{\bm\phi}^1\Psi^{Ds}_{m_\ell}({\bf P};{\bf k}_1,{\bf k}_2)
\eea
where $m_\ell$ is the projection of the spin-2 spatial wave function in some (arbitrary) fixed direction, and the ${\bm\phi}^{0,1}$ are the isospin operators (\ref{eqFlavor}). 
In parallel with (\ref{eq:4.15}), the spatial wave functions will be written
\bea
\Psi_{m_\ell}^{Da}=&& \sqrt{\sfrac{4\pi}2} \Big[
{\bf k}_1^2Y^2_{2m_\ell}(\hat {\bf k}_1)-{\bf k}_2^2Y^2_{m_\ell}(\hat {\bf k}_2)\Big]\Phi_D
\nonumber\\
\Psi_{m_\ell}^{Ds}= &&\sqrt{4\pi}\Big[\cos\phi \,{\bf k}^2
Y^2_{m_\ell}(\hat{\bf k})+\sin\phi\, {\bf r}^2 Y^2_{m_\ell}(\hat{\bf r})\Big]\Phi_D\qquad
\eea   
where $\Psi^{Da}$ is the most general $L=2$ function depending on 
${\bf k}_1$ and ${\bf k}_2$ that is antisymmetric under the interchange 
${\bf k}_1\leftrightarrow {\bf k}_2$, and $\Psi^{Ds}$ 
the most general symmetric function, and $\Phi_D$ is a symmetric D-state counterpart to the S-state function $\Phi_S$.  Note that the mixing between the two terms in $\Psi^{Da}$ is fixed by the antisymmetry requirement, but that the requirement of symmetry alone cannot fix the relative contributions of the two terms in $\Psi^{Ds}$.  We will fix these below.

To complete the construction of the D-state wave function we combine the $L=2$ orbital part with the {\it total\/} quark spin 3/2 wave functions.  For a spin-1/2 state with projection $\lambda$ this combination is
\bea
\Psi_{\lambda}^{D\eta}&=&\sum_{m_\ell=-1}^2\left<2\,m_\ell\,\sfrac32\, \mu | \sfrac12\,\lambda\right>\Psi_{m_\ell}^{D\eta} \left| \sfrac32\, \mu\right>,\qquad\quad \label{eq2.13}
\eea
where $\eta=\{a,s\}$, and the value of $\mu$ is fixed by the vector coupling coefficient, and the spin 3/2 state can be written as a direct product of a vector and a spin 1/2 spinor:
\bea
\left| \sfrac32\, \mu\right>=\sum_{\nu}\left<1\,\Lambda\,\sfrac12\,\nu | \sfrac32\,\mu\right> \left|1\,\Lambda\right>\otimes \left| \sfrac12\,\nu\right>\qquad \label{eq:114}
\eea
where $\left|1\,\Lambda\right>=\varepsilon^i_{\Lambda}$ is the spin function of the (12) pair (spin 1 and polarization $\Lambda$), and $\left| \sfrac12\,\nu\right>$ is the spin of the spectator quark 3. It is convenient to project out the polarization state of the diquark, and the wave function we will use in applications will be constructed from the functions
\begin{align}
\Theta_{\Lambda \lambda}^D&({\bf k}_i)\equiv\sqrt{4\pi}
\sum_{m_\ell=-1}^2\left<2\,m_\ell\,\sfrac32\, \mu | 
\sfrac12\,\lambda\right> {\bf k}_i^2 Y_{2m_\ell}(\hat {\bf k}_i) 
\left<1\,\Lambda| \sfrac32\, \mu\right>
\nonumber\\
&=\sqrt{4\pi}\sum_{m_\ell=-1}^2\left<2\,m_\ell\,\sfrac32\, \mu | 
\sfrac12\,\lambda\right> {\bf k}_i^2Y_{2m_\ell}(\hat {\bf k}_i) 
\nonumber\\
&\qquad\qquad\qquad\times
\left<1\,\Lambda\,\sfrac12\,\nu | \sfrac32\,\mu\right> \left| \sfrac12\,\nu\right>
.\label{eq2.13a}
\end{align}  
The complete three-quark nonrelativistic symmetric D-state is then
\begin{align}
&\Psi^{Da}_{\Lambda \lambda}({\bf P};{\bf k}_1,{\bf k}_2)=\frac{1}{\sqrt{2}} \Big(\Theta^D_{\Lambda \lambda}({\bf k}_1)-\Theta^D_{\Lambda \lambda}({\bf k}_2)\Big)\,\Phi_D
\nonumber\\
&\Psi^{Ds}_{\Lambda \lambda}({\bf P};{\bf k}_1,{\bf k}_2)=\Big(\cos\phi\,\Theta^D_{\Lambda \lambda}({\bf k})+\sin\phi\,\Theta^D_{\Lambda \lambda}({\bf r}) \Big)\,\Phi_D,
\label{eq:127a}
\end{align}
Note that both of these components 
depend on the diquark (spin-1) polarization $\Lambda$. 

Before moving on with our construction, we call attention to the work of Diaz and Riska \cite{DiazRiska}, who present a construction of a D-state wave function along lines similar to ours.  However, we were unable to confirm their results.

\subsubsection{Alternative nonrelativistic form}

To prepare for what follows, it is useful to rewrite the $\Theta$ functions in Eq.~(\ref{eq:127a}) in an alternative form.  In Appendix  \ref{apTheta}, we show that 
\bea
\Theta_{\Lambda\lambda}^{D}({\bf k}_i)&=& 
\sqrt{\sfrac{3}{2}}\,
({\bm \varepsilon}^*_{\Lambda} )_\ell
 D^{\ell \ell'}({\bf k}_i)\,\sigma_{\ell'} \left|\sfrac12\lambda\right>,
 \label{eq:119}
\eea   
where 
$\varepsilon_{\Lambda}$ is a spin-1 polarization vector of the (12) 
pair with spin projection $\Lambda$ (already introduced above), 
and $\left|\frac12\lambda\right>$ is the two-component 
spinor with spin projection $\lambda$, both along 
the $\hat z$ direction, and the angular momentum two $D$ matrix is
\bea
D^{\ell m}({\bf k}_i)={\bf k}_i^\ell {\bf k}_i^m-\sfrac13\,
{\bf k_i}^2\,\delta_{\ell m}. 
\label{eq:120}
\eea 

\subsubsection{Normalization}

The wave function (\ref{eq:127a}) depends on the momenta of all three of the quarks.  Extracting the diquark content requires that we average over the internal momenta of the 12 pair (the diquark), 
just as we did in Eq.~(\ref{eq:4.17}).  

Using the new representations for the $\Theta$'s, it is now possible 
to separate the internal diquark variable ($r$) from the quark variable ($k$).  
Denoting 
${\bf k}_1\equiv {\bf k}_+=\frac12{\bf k}+{\bf r}$ and 
${\bf k}_2\equiv {\bf k}_-=\frac12{\bf k}-{\bf r}$,  
 the $D$ of Eq.~(\ref{eq:120}) separates
\bea
D^{\ell m}({\bf k}_\pm)&=&\sfrac14 D^{\ell m}({\bf k})+D^{\ell m}({\bf r})\pm\sfrac12 G^{\ell m}({\bf k},{\bf r})\qquad
 \label{eq:132}
\eea   
with 
\bea
G^{\ell m}({\bf k},{\bf r})\equiv {\bf k}^\ell {\bf r}^m+{\bf r}^\ell {\bf k}^m -\sfrac23 \delta_{\ell m} {\bf k}\cdot {\bf r}
\eea

With this substitution, the antisymmetric component becomes
\bea
\Psi^{Da}_{\Lambda\lambda}({\bf P};{\bf k}_1,{\bf k}_2) =&&\sqrt{\sfrac34}\,
({\bm \varepsilon}_\Lambda^*)_\ell\, G^{\ell m}({\bf k},{\bf r})
\nonumber\\&&\qquad\times 
\sigma_m \,\left|\sfrac12\,\lambda\right> 
\Phi_D({\bf P};{\bf k}_1,{\bf k}_2).\qquad\label{eqPsi0a}
\eea    
Note that this component is linear in both
${\bf r}$ and ${\bf k}$,  describing the coupling of a diquark with an {\it internal P-wave, or internal angular momentum\/}-1 structure to the third quark in a relative P-wave, while the isovector component (\ref{eq:127a})  is a sum of two D-wave components, one in the total diquark momentum 
${\bf k}$ and the other in the {\it internal\/} diquark momentum ${\bf r}$.   

In matrix elements, when we average over the direction of 
the relative momentum, ${\bf r}$,  
these components will give a non-zero result only when multiplied by another component of the same type.  Using the results from Appendix \ref{apNorma},  and the isospin matrix elements (\ref{eq:4.16}), the normalization integral becomes
\bea
Q=&& e_0\Big[\sfrac12+\sfrac32\tau_3\Big] \cos^2\!\theta_D\,\left<\left|\Psi^{Da}\right|^2\right>_{\lambda'\lambda} 
\nonumber\\&&
+ e_0 \Big[\sfrac12-\sfrac12\tau_3 \Big]  \sin^2\!\theta_D\,\left<\left|\Psi^{Ds}\right|^2\right>_{\lambda'\lambda}  
\nonumber\\
=&&\sfrac12 e_0\Big[1 +\tau_3\left(3  \cos^2\!\theta_D- \sin^2\!\theta_D\right)\Big]\,\delta_{\lambda'\lambda}\,{\cal N}_D\qquad
\eea  
with the normalization constant ${\cal N}_D$ defined in Eqs.~(\ref{eq:134}) and (\ref{eq:C7}).  Once again we observe that the correct neutron charge cannot be obtained unless $\theta_D=\pi/4$, implying an equal mixture of diquark  symmetric and antisymmetric components, just as for the S-state.  The D-state wave function is then normalized to ${\cal N}_D=1/e_0$.

\subsubsection{Defining the nonrelativistic quark-diquark wave function}

The D-state presents us with some new issues in extracting a  quark-diquark wave function from the full three-quark wave function.  Examination of Eqs.~(\ref{eq:127a}) and (\ref{eqPsi0a}) shows that there are three orthogonal structures, each with a different  kind of average over the diquark internal momentum variable ${\bf r}$.  The first of these accompanies the square of the  $\Theta^D_{\Lambda \lambda}({\bf k})$ term in the isovector term.  Including an extra factor of 1/4 (to simplify the final normalization) gives a definition 
\bea
\frac14\int_{k,r}^{\rm NR} {\bf k}^4\,\Phi_D^2=\int\frac{d^3 k}{(2\pi)^3}\, 
{\bf k}^4\,\phi^2_D
\label{eq:4.38}
\eea  
where the integral was defined in Eq.~(\ref{eq:2.12}), and to simplify the notation we let $\Phi_D({\bf P}; {\bf k}_1, {\bf k}_2)\to \Phi_D$, and the new wave function $\phi_D=\phi_D({\bf P}, {\bf k})$.  Except for the factor of 1/4, this is precisely the same prescription used for the S-state wave function [see Eq.~(\ref {eq:4.17}) for the arguments and other details], and leads to the replacement 
\bea
\sfrac12\Phi_D\to  \phi_D .
\eea
This function is normalized to
\bea
\int\frac{d^3 k}{(2\pi)^3}\,{\bf k}^4\phi^2_D=
{\cal N}_D=\frac{1}{e_0}, \label{eq:Dnorm}
\eea  
which, because of the definition (\ref{eq:4.38}), is equivalent to the normalization defined in Eqs.~(\ref{eq:134}) and (\ref{eq:C7}).

The second term to be defined depends linearly on ${\bf r}$ and appears only in the isoscalar term (\ref{eqPsi0a}).    As previously mentioned, this linear ${\bf r}$ dependence describes a diquark with an internal {\it angular momentum\/} dependent P-wave, which can be represented by a diquark with a polarization vector ${\bm \zeta}_\nu$ (with $\nu=\{\pm,0\}$ the three independent polarization states).  But this isoscalar diquark is described by {\it two\/} vectors: one due to its internal momentum, ${\bm \zeta}_\nu$, and the other due to its  {\it spin\/}, ${\bm \varepsilon}_\Lambda$.   This term is orthogonal to all other terms, and the diquark content of this term can be extracted by introducing the following correspondence 

\bea
\int_{k,r}^{\rm NR} {\bf k}^2\,{\bf r}^\ell{\bf r}^m\, 
\Phi_D^2 &&=  \delta_{\ell m}\frac13 \int_{k,r}^{\rm NR} \,{\bf k}^2{\bf r}^2\, 
\Phi^2_D
\nonumber\\
&&=\delta_{\ell m}\frac{c_P^2}3  \int_{k,r}^{\rm NR} \,{\bf k}^4\, \Phi^2_D
\nonumber\\
&&\to\frac{4c_P^2}3  \sum_\nu \zeta_{\nu}^{\ell}\zeta_{\nu}^{m *}
\int \frac{d^3k}{(2\pi)^3} \, {\bf k}^4\,\phi_D^2
, \qquad\qquad \label{eq:136}
\eea 
where the factor $c^2_P=1/A=3/20$ was computed in Appendix \ref{apNorma} [see Eq.~(\ref{eq:C7})] and the extra factor of 4 from the correspondence (\ref{eq:4.38}) has been included.
Note that the integral over the continuous variable ${\bf r}$ is replaced by the sum of the complete set of polarization states ${\bm \zeta}_\nu$, 
and the average over 
${\bf r}^2$ is replaced by $c_P^2 \,{\bf k}^2$.  
With this choice we obtain the correspondence

\bea
\sfrac12{\bf r}^\ell \Phi_D\to \frac{c_P}{\sqrt{3}}\, |{\bf k}| 
\,{\bm \zeta}_\nu^\ell \,\phi_D \label{eq:4.42}
\eea    
 with the sum over polarization states, $\nu$, 
to be carried out in the calculation of any matrix element, 
using the completeness relation
 \bea
\sum_\nu{\bm \zeta}_{\nu}^\ell {\bm \zeta}_{\nu}^{m\,*}&&=\delta_{\ell m} \, , \label{eq:4.43}
 \eea
and with the understanding that  the terms linear in the total diquark momentum, ${\bf k}$, will be left unchanged.  
The reason for using the momentum ${\bf k}$ (instead of ${\bf r}$) 
on the r.h.s.~of~(\ref{eq:4.42}) is to eliminate all dependence on 
${\bf r}$,  leaving only one function $\phi_D$ 
with the same normalization condition (\ref{eq:Dnorm}).

Finally, the last term to be defined is the term depending on $\Theta^D_{\Lambda \lambda}({\bf r})$.   This term is a contraction of the {\it spin\/}-1 vector, ${\bm \varepsilon}_\Lambda$, of the diquark with the D-wave internal {\it momentum\/} structure of the diquark.  Its average can be represented by a new effective diquark polarization vector, ${\bm \epsilon}_{D\Lambda}$,  with the familiar property (\ref{eq:4.43}).    The average we need is
\bea
&&\sum_\Lambda\int_{k,r}^{\rm NR}({\bm \varepsilon}_\Lambda)_{\ell'} D^{\ell' m'}({\bf r}) ({\bm \varepsilon}^*_\Lambda)_\ell D^{\ell m}({\bf r})\Phi_D^2 
\nonumber\\&&
= \delta_{m' m}  \frac29 \int_{k,r}^{\rm NR} {\bf r}^4\, \Phi^2_D
=\delta_{m' m}\frac{2c_D^2}9  \int_{k,r}^{\rm NR} \,{\bf k}^4\, \Phi^2_D
\nonumber\\
&&\to\frac{8c_D^2}9  \sum_\Lambda {\bm \epsilon}_{D\Lambda}^{m'}
{\bm \epsilon}_{D\Lambda}^{m *}\int \frac{d^3k}{(2\pi)^3} \, {\bf k}^4\,\phi_D^2
, \qquad\qquad \label{eq:4.44}  
\eea  
where $c_D^2=B/A=1/16$ was computed in Appendix \ref{apNorma}, and the factor of 4 from (\ref{eq:4.38}) has again been included.  With this choice we obtain the correspondence
\bea
 \sfrac12({\bm \varepsilon}^*_\Lambda)_\ell D^{\ell m}({\bf r})
\Phi_D\to\frac{\sqrt{2}c_D}{3} {\bm \epsilon}_{D\Lambda}^{m *}\,{\bf k}^2
\phi_D\, .
\eea

With this notation, the quark-diquark wave functions corresponding 
to the three-quark wave functions given in Eqs.~(\ref{eq:127a}) 
and (\ref{eqPsi0a}) become
\begin{align}
\psi^{Da}_{\Lambda \lambda}(P,k)=&\,
c_P 
({\bm \varepsilon}^*_\Lambda)_\ell\,G^{\ell m}({\bf k},{\bm \zeta}_\nu)
\sigma_{m}\left|\sfrac12\,\lambda\right>\,|{\bf k}|\,\phi_D(P,k) 
\nonumber\\
\psi^{Ds}_{\Lambda\lambda}(P,k)=&\frac2{\sqrt{5}}\;\Theta_{\Lambda\lambda}^D({\bf k})\phi_D(P,k)
\nonumber\\
+&\quad\frac{4c_D}{\sqrt{15}} \,{\bm \epsilon}_{D\Lambda}^{m *}
\sigma_m\left|\sfrac12 \lambda\right>\, {\bf k}^2 
\phi_D(P,k),\label{eqPsi1b} 
\end{align}       
where the value of the mixing angle $\phi$ determined in Appendix \ref{apNorma} has been used.

It is straightforward to confirm that our substitutions preserve the normalizations given in Appendix \ref{apNorma}.  We now turn to the relativistic generalization  of Eq.~(\ref{eqPsi1b}).

\subsubsection{Relativistic quark-diquark D-state wave function}


Using the ideas developed in Ref.~\cite{NDeltaD,DeltaDFF}, the relativistic 
analogue of the wave functions from the previous section 
are constructed from the four-vector 
\bea
\tilde k=k-\frac{(k\cdot P)}{M^2}P,
\label{eq:ktilde}
\eea
where $P$  is the total four momentum of the nucleon.   
This insures that $\tilde k$ will reduce to ${\bf k}$ in the rest system. 
Similarly, all of the polarization four-vectors 
are chosen to be orthogonal to $P$, insuring that, 
in the nucleon rest frame,  their time components 
are zero and their spatial components 
are identical to their corresponding non-relativistic 
three-vectors. In addition, we replace
\bea
\delta_{\ell\ell'}&\to& -\tilde g_{\alpha\beta} =-\Big(g_{\alpha\beta}-\frac{P_\alpha P_\beta}{M^2}\Big)
\nonumber\\
({\bm \varepsilon}_\Lambda^\ast)_{\ell}  &\to& 
(\varepsilon_\Lambda^\ast)^\alpha 
\nonumber\\
\sigma_{\ell'}\left|\sfrac12\,\lambda\right>&\to&\sqrt{3}\;U^\beta(P,\lambda)
\nonumber\\
G^{\ell\ell'}( {\bf k}, \zeta_\nu )  &\to&  
G^{\alpha\beta}(k,\zeta_\nu ) 
\equiv 
\tilde k^\alpha \zeta_\nu^\beta 
+ \zeta_\nu^\alpha \tilde k^\beta -
\sfrac23\tilde g^{\alpha\beta}\,(\tilde k \cdot \zeta_\nu)  
\nonumber\\
D^{\ell\ell'}({\bf k})&\to& D^{\alpha\beta}(P,k)\equiv \tilde k^\alpha \tilde k^\beta-\sfrac13\tilde g^{\alpha\beta}\,\tilde k^2 \qquad 
\nonumber\\
\phi_D(P,k)&\to&\psi_D(P,k),  \label{eq2.47}
\eea
where $\ell\to\alpha$, $\ell'\to\beta$
and $U^\beta(P,k)$ was defined in Eq.~(\ref{eq2.8}).  
With these correspondences the total D-state wave function of the nucleon for an on-shell diquark (composed of quarks 1 and 2) and an off-shell quark (3) with momentum $k_3=P-k$ is, 
\bea
\Psi^{D}_{\Lambda\lambda}(P,k )=\frac{1}{\sqrt{2}} \Big\{{\bm\phi}^0\psi^{Da}_{\Lambda\lambda}(P,k)+{\bm\phi}^1\psi^{Ds}_{\Lambda\lambda}(P,k)\Big\}\qquad \label{eq:4.51}
\eea
where
\begin{align}
\psi^{Da}_{\Lambda \lambda}(P,k)=&
\sqrt{3}\,c_P ({\varepsilon}^*_\Lambda)_\alpha
G^{\alpha\beta}(\tilde k, \zeta_\nu)
U_\beta(P,\lambda)|\tilde k|\psi_D(P,k) \qquad
\nonumber\\
\psi^{Ds}_{\Lambda\lambda}(P,k)=&\frac2{\sqrt{5}}\;\Theta_{\Lambda\lambda}^D(P,k)\psi_D(P,k)
\nonumber\\
&\quad-\frac{4\,c_D}{\sqrt{5}} \,{\epsilon}_{D\Lambda}^{\beta *}U_\beta(P,\lambda)\,\tilde k^2\psi_D(P,k),\label{eq1.18a} 
\end{align}       
where $G^{\alpha\beta}(\tilde k, \zeta_\nu)$ is the straightforward generalization of its nonrelativistic counterpart, $|\tilde k|\equiv\sqrt{-\tilde k^2}$,
\bea
\Theta_{\Lambda\lambda}^D(P,k)=
\frac3{\sqrt{2}}
(\varepsilon_\Lambda^\ast)_\alpha   D^{\alpha\beta}(P,k) U_\beta(P,\lambda),
\eea
and $\phi_D(P, k)$ is a spherically symmetric scalar function of the off-shell quark momenta $k_3=P-k$.

\subsubsection{Normalization of the relativistic wave function}

The normalization of the relativistic 
D-state wave function, like the S-state, 
is 
fixed by the charge (\ref{eqCharge}):
\bea
Q=
3\sum_\Lambda\int_{k} \overline{\Psi}_{\Lambda\lambda^\prime}^{D}(P,k) 
j_q(0) \gamma^0 \Psi_{\Lambda\lambda}^D(P,k).
\label{eqChargeN}
\eea
%
This condition is satisfied by the wave function (\ref{eq:4.51}), with components (\ref{eq1.18a}), provided
\bea
1=e_o\int \frac{d^3 k}{(2\pi)^3 2E_k} \tilde k^4\psi^2_D(P,k)\, , \label{eq:normD1}
\eea
which is the relativistic analogue of the nonrelativistic norm (\ref{eq:Dnorm}).

From now on we will use $E_k= \sqrt{m_s^2 + {\bf k}^2}$ 
instead of $E_s$.

\subsection{P-state component}

 The origin of the P-state component can be traced back to the general CST relation between the relativistic vertex function, $\Gamma$, and the relativistic wave function 
 \bea
 \Psi(P,k)=\frac{1}{m_q-\slashed{k}_3}\,\Gamma(P,k)=\frac{m_q+\slashed{k}_3}{m^2_q-k_3^2}\,\Gamma(P,k)\qquad
 \eea
 where $(m_q-\slashed{k}_3)^{-1}$ is the propagator of the off-shell quark with four-momentum $k_3=P-k$ (when quarks 1 and 2 are the on-shell spectators).  In the presence of confinement, the pole at $m_q^2=k_3^2$ is cancelled by a zero $\Gamma$, so the wave function can be modeled without regard to this singularity.  For simplicity, in our previous work \cite{Nucleon} we also absorbed the projection operator  $m_q+\slashed{k}_3$ into our model of $\Psi$, but the assumption that the resulting $\Psi$ is a {\it pure\/} S-state ignores some additional structure that the projection operator  could provide.   
 
Now, suppose we assume that the projection operator is {\it not\/} absorbed into the definition of the wave function.  Then the wave function would be
\bea
\Psi_0^S(P,k)\simeq \frac{\Gamma(P,k)}{m_q^2-k_3^2}
\eea
and the action of the quark projection operator on this wave function would give
 \bea
 (m_q+\slashed{k}_3)\,\Psi_0^S(P,k)&&=(m_q+M-\slashed{k})\,\Psi_0^S(P,k)
 \nonumber\\
 &&=\left[m_q+M-\frac{P\cdot k}{M}-\slashed{\tilde k}\right]\Psi_0^S(P,k)
 \nonumber\\
 &&\simeq\Psi^S(P,k)+n_P\,\slashed{\tilde k}\,\Psi^P(P,k),
 \label{eq:Porigin}
 \eea
where $\tilde k$ was defined by Eq.~(\ref{eq:ktilde}),  and the expressions were reduced using the Dirac equation, $\slashed{P} \Psi =M\Psi$, satisfied by the wave function.  This clearly shows how the Dirac  projection operator leads both to a redefinition of the S-state and to a new P-state component.  By allowing the mixing parameter, $n_P$, to be determined by the data, we free ourselves from any biases about the possible size of this effect.

A P-state component could have been constructed by starting from either the S or D-state contributions, but in this paper we limit the discussion to P-states constructed from the dominant S-state component, as suggested by (\ref{eq:Porigin}).  This gives the following ansatz for the P-state
\ba
\Psi_{\Lambda \lambda}^P(P,k)&=&
\frac{1}{\sqrt{2}} \,\slashed{\tilde k} 
\left[{\bm\phi}^0 u(P,\lambda) -
{\bm\phi}^1 \varepsilon^{\alpha\,\ast }_{\Lambda P}\,U_\alpha(P,\lambda) 
\right] \nonumber \\
& &
\qquad \times\psi_P(P,k). \label{eq:4.62}
\ea
The normalization of the P-state is
\ba
e_0 \int \frac{d^3k}{(2\pi)^32E_k}\,(-\tilde k^2)\, \psi^2_P(P,k)=1,
\ea
where the minus sign arrises because $\slashed{\tilde k}\gamma^0\slashed{\tilde k}=-\tilde k^2>0$.

Our S-state wave function is a Dirac spinor with the {\it lower\/} two components exactly zero in the nucleon rest frame.  Conversely, the wave function (\ref{eq:4.62}) is a Dirac spinor with {\it upper\/} two components exactly zero in the nucleon rest frame.   For this reason it has positive parity (the negative parity of a P-wave being cancelled by the negative sign from the Dirac parity operator $\gamma^0$).  

 This P-wave component is of purely relativistic origin because it will only make nonzero contributions to matrix elements which have lower, relativistic components, in Dirac space.

\section{Breaking isospin symmetry } \label{udmix}

In fitting the DIS cross sections, it will be necessary to include the possibility that the $u$ and the $d$ distributions are {\it not\/} identical.   This assumption violates isospin invariance.  

In the following discussion, the ``$u$ distribution'' refers to the $u$ distribution in protons and the $d$ distribution in neutrons (we retain charge symmetry).  Similarly. the ``$d$ distribution'' refers to to the $d$ distribution in protons and the $u$ distribution in neutrons. 

To generalize the formalism to allow for this difference, look at the flavor wave functions (\ref{eqFlavor}) in more detail.  Examination of their structure shows that some of the flavor functions describe $u$ distributions and others $d$ distributions. 
In particular, the isoscalar and the $\ell=0$ isovector components describe $u$ quark distributions, since the interacting quark is always a $u$ quark in the proton or a $d$ quark in the neutron.   Returning to the full three-quark notation of Ref.~\cite{Nucleon}, where the flavor wave functions were written as a direct product in the order 1,2,3, so that  $a_1b_2c_3\to abc$, and identifying the (12) pair with the diquark, so that the photon interacts with the third quark, it is easy to extract the separate $u$ and $d$ distributions
\bea
\sqrt{\sfrac12}(ud-du)\begin{cases} u  & {\rm proton}\cr d & {\rm neutron} \end{cases}&&\to {\bm\phi}^0\chi^t
\nonumber\\
-\sqrt{\sfrac16}(ud+du) \begin{cases} u  & {\rm proton}\cr d & {\rm neutron} \end{cases}&& \to {\bm\phi}^1_{\ell=0}\chi^t
\eea
while the isovector $\ell\ne0$ components describe the $d$ quark distribution in the proton and the $u$ quark distribution in the neutron:
\bea
\sqrt{\sfrac23}\begin{cases} \phantom{-}(uu) d 
&  {\rm proton}\cr -(dd)u  
& {\rm neutron} \end{cases} \to {\bm\phi}^1_{|\ell|=1}\chi^t\, .
\eea

Next, we introduce different wave functions for the  $u$ and $d$ quarks for each angular momentum component
\bea
\psi_L(P,k)\to \begin{cases} \psi^{L}_u(P,k) & u\quad {\rm quark}\cr
\psi^{L}_d(P,k) & d\quad{\rm quark}
\end{cases}
\eea
where $L=$S, P, D.  We assume that (for example) the $u$ quark distribution corresponding to the spin-0 and spin-1 diquarks are identical, but this could be relaxed later, if necessary.   Then, the matrix element of the charge accompanying the isoscalar diquark is unchanged except for replacing $\psi^2_L(P,k)$ by $[\psi^{L}_{u}(P,k)]^2$.   However, the isovector matix element, $j^S\,\psi^2_L$ [recall Eq.~(\ref{eq:4.16})], now separates into two independent contributions.   
Allowing for the fact that the renormalization
of the $u$ quark charge ($e_0\to e^0_u$) and $d$ quark charge  ($e_0\to e^0_d$) might be unequal, and dropping the index $L$, gives
\bea
j^S_0\psi^2_u(P,k)&=&(\tau\cdot\xi_0)(\sfrac16+\sfrac12 \tau_3) (\tau\cdot\xi^*_0) \,e^0_u\psi_u^2(P,k)
\nonumber\\
&=&(\sfrac16+\sfrac12 \tau_3)\,e^0_u \psi_u^2(P,k)
\nonumber\\
j^S_1\psi_d^2(P,k)&=&(\tau\cdot\xi_+)(\sfrac16+\sfrac12 \tau_3) (\tau\cdot\xi^*_+)\,e^0_d\psi_d^2(P,k)
\nonumber\\
&&+(\tau\cdot\xi_-)(\sfrac16+\sfrac12 \tau_3) (\tau\cdot\xi^*_-)\,e^0_d\psi_d^2(P,k)
\nonumber\\
&=&(\sfrac13- \tau_3)\,e^0_d\psi_d^2(P,k)\, .
\eea
Hence the generalized isovector charge operator is transformed (for any $L$) to
\bea
j^S\psi^2(P,k)\to &&j^S_0\psi_u^2(P,k)+j^S_1\psi_d^2(P,k)
\nonumber\\
=&&\sfrac16(1-\tau_3)\Big[ e^0_u\psi_u^2(P,k)+2\, e^0_d\psi_d^2(P,k)\Big]
\nonumber\\
&& +\sfrac23\tau_3\Big[\, e^0_u\psi_u^2(P,k) -\, e^0_d\psi_d^2(P,k)\Big]\label{eq:2.24}
\eea
If $\psi_d=\psi_u$  (and $e^0_u=e^0_d$) then the sum of these reduces to the previous result (\ref{eq:4.16}), but $\psi_d\ne\psi_u$  the last term clearly gives a different result.  When combined with the isoscalar contribution, the total is (choosing the mixing angle $\theta_L=\pi/4$ as before)
\bea
&&\sfrac12 j^A\psi_u^2(P,k)+\sfrac12 \left[j_0^S\psi_u^2(P,k)+j_1^S\psi_d^2(P,k)\right]
\nonumber\\
&& =
\sfrac16 (2 e^0_u\psi_u^2 +  e^0_d\psi_d^2) +
\sfrac12 (2 e^0_u\psi_u^2 -  e^0_d\psi_d^2) \tau_3
\nonumber \\
&&\quad=\begin{cases}  
\phantom{-}
\sfrac13 (4 e^0_u\psi_u^2 -  e^0_d\psi_d^2) & p \cr 
-\sfrac23 ( e^0_u\psi_u^2 -  e^0_d\psi_d^2)   & n\, . \end{cases}
\eea 
This gives the requirement that both $u$ and $d$ distributions be normalized to $1/e^0_q$.  For arbitrary $L$ this gives
\bea
1&&=e^0_u\int \frac{d^3 k}{(2\pi)^3 2E_k}(- \tilde k^2)^L[\psi^L_u(P,k)]^2
\nonumber\\
&&=e^0_d\int \frac{d^3 k}{(2\pi)^3 2E_k} (-\tilde k^2)^L[\psi^L_d(P,k)]^2\, .
\eea

This result for the charge operator may be generalized.  In particular, if ${O}$ is any operator in isospin space, and the isospin one matrix elements of $O$ are defined by 
\bea
&&3\,{\bm \phi}^1_0 O {\bm \phi}^1_0\equiv O^1_0
\nonumber\\
&&3\,{\bm \phi}^1_1 O {\bm \phi}^1_1+3\,{\bm \phi}^1_{-1} O {\bm \phi}^1_{-1}\equiv O^1_1
\eea
then for arbitrary angular momentum components $L$ and $L'$, we may make the replacement
\bea
\left<{O}\right>_{L', L}\to&&(O+O^1_0) \,\psi_u^{L'}(P,k) \psi_u^L(P,k) 
\nonumber\\
&&+O^1_1\,\psi_d^{L'}(P,k)\psi_d^L(P,k). \qquad
\eea

\section{Summary and overview}

In this paper we present relativistic CST wave function for the nucleon with S, P, and D-state components.  These wave functions are designed to be used in calculations where two of the quarks are non-interacting on-shell spectators, with the third off-shell quark interacting with an external probe.  In such a situation the full dependence of the matrix element on the relative momentum, $r$, of the two on-shell spectators is contained in the three-quark wave function.   Integrating over $r$ will then lead to a new effective wave function with the two non-interacting quarks replaced by a quark pair (an effective diquark) with a mass, $m_s$ which may be treated as a parameter.  The resulting quark-diquark wave function contains all of the information originally included in the three-quark wave function and may be used in a variety of calculations.

The extraction of a quark-diquark wave function is discussed in general terms in Sec.~\ref{sec:IIB} with the most general result for the extraction of the wave function  presented in Eq.~(\ref{eq:2.8}).  In Sec.~\ref{sec:IV} the various component wave functions are constructed and presented.  The relativistic S-state quark-diquark wave function is given in Eq.~(\ref{eqPsiN}).  This is identical to the S-state wave function  used previously \cite{Nucleon,NDeltaD}, but the extraction of this wave function from the full three-quark wave function, culminating in the nonrelativistic correspondence (\ref{eq:4.17}), has never been discussed before.   The relativistic P-state quark-diquark wave function, also new, is given in Eq.~(\ref{eq:4.62}).   This component vanishes in the nonrelativistic limit.

The bulk of this paper is devoted to obtaining the relativistic D-state quark-diquark wave function given in Eqs.~(\ref{eq:4.51})-(\ref{eq1.18a}).  This component is a superposition of terms with an isoscalar P-wave diquark (represented by the polarization vector $\zeta_\nu$), and  two terms with an isovector diquark, one with an internal S-wave structure and one with an internal D-wave structure.  Although each of these diquarks has a different angular momentum structure, they are all derived from contributions in which the quarks are in a relative $L=2$ angular momentum state.  Furthermore, each of these diquarks has spin-1, as required by the coupling of overall spin 3/2 to $L=2$ to produce a nucleon with spin-1/2.   These three components are orthogonal, and the {\it only\/} component that can interfere with the dominant S-state component is the latter, 
which is summarized by the contribution
\bea
\Psi^{D}_{\Lambda\lambda}(P,k )\to&&\,\frac{3}{\sqrt{5}}{\bm\phi}^1\,
(\varepsilon_\Lambda^\ast)_\alpha  D^{\alpha\beta}(P,k) U_\beta(P,\lambda)
\psi_D(P,k) 
\nonumber\\
\equiv&& \;\Psi^{D,2}_{\Lambda\lambda}(P,k)\, .\label{dmain}
\eea
In applications when we need terms linear in the D-state,  this is the only term that need be considered, and it can contribute only to matrix elements which do not allow for a complete integration over all directions of ${\bf k}$ (as in the important case of DIS). 

The normalization condition for the component (\ref{dmain}) is
\bea
3\sum_\Lambda&&\int_{k} \overline{\Psi}_{\Lambda\lambda^\prime}^{D,2}(P,k) 
j_q(0) \gamma^0 \Psi_{\Lambda\lambda}^{D,2}(P,k)
\nonumber\\
&&=j^S\delta_{\lambda'\lambda} \,\frac2{5}
\eea
where $j^S$ was defined in Eq.~(\ref{eq:4.16}) and the D-state normalization condition (\ref{eq:normD1}) has been retained.  Recalling that the contribution from both the isovector D-state components is 1/2, we recover the previous result, also contained in (\ref{eq1.18a}), that the component (\ref{dmain}) accounts for 4/5 of the total symmetric contribution.

To allow for the possibility that the $u$ and $d$ quarks have a different angular momentum distribution, we may break isospin symmetry as discussed in Sec.~\ref{udmix}.   The symmetry can be broken and the shapes of the $u$ and $d$ quark distributions individually adjusted as long as their normalization remains fixed.

The wave functions derived in this paper are used in our discussion of DIS in the companion paper \cite{following}, which provides numerical calculations indicating
that the $L\ne 0$ wave function components described here make important contributions to the proton spin.  

\acknowledgements

This work was partially support by Jefferson Science Associates, 
LLC under U.S. DOE Contract No. DE-AC05-06OR23177.
This work was also partially financed by the European Union
(HadronPhysics2 project ``Study of strongly interacting matter'')
and by the Funda\c{c}\~ao para a Ciencia e a
Tecnologia, under Grant No.~PTDC/FIS/113940/2009,
``Hadron structure with relativistic models''.  
G.~R.~was supported by the Portuguese Funda\c{c}\~ao para 
a Ci\^encia e Tecnologia (FCT) under Grant  
No.~SFRH/BPD/26886/2006.






\appendix

\section{Relativistic diquark averages}
\label{apDiquark}
 
In this Appendix we discuss the reduction of the integral (\ref{eq1.29a}) to the form (\ref{eq1.30}).  Begin by writing the integral in a manifestly covariant form and replacing the integrals over ${k}_1$ and ${k}_2$ by integrals over the sums and differences, $k$ and $r$, of these momenta 
\begin{align}
(2\pi)^6 \int_{k_1k_2}=&\int d^4k_1 d^4k_2\; 
\delta_+(m_q^2-k_1^2)\delta_+(m_q^2-k_2^2)
\nonumber\\
=&\int d^4k \int d^4 r\; \delta_+\Big(E_1^2-(\sfrac12k_0+r_0)^2\Big)
\nonumber\\
&\qquad\times\delta_+\Big(E_2^2-(\sfrac12k_0-r_0)^2\Big) 
\end{align}
Both integrals are covariant, so we will evaluate the integral over $r$ in the {\it two body rest system\/}, and use the two $\delta$ functions to fix $r_0$ and $|{\bf r}|$.  This will leave the integral over $k$ completely unconstrained, except we {\it define\/} $k^2=s$.  Therefore, in the two body rest system $k_0=\sqrt{s}$ and the integral is  
\begin{align}
(2\pi)^6 \int_{k_1k_2}=\sfrac14\int d^4k \int d \Omega_{\hat {\bf r}}\; \sqrt{\frac{s-4m_q^2}{4s}}
\end{align}
Finally we express the $k_0$ integration in terms of $s$ using $k_0^2=s+|{\bf k}|^2\equiv E_s^2$, giving the relation (\ref{eq1.30})
\bea
 \int_{k_1k_2}=\int_{s\,k}\, .
\eea
In using this formula, we must be careful to evaluate angular integrals involving ${\bf r}$ in the rest system of the two-body pair.

\section{Alternative form for the nonrelativistic D-wave}
\label{apTheta}

To show the equivalence of (\ref{eq2.13a}) and (\ref{eq:119}), first introduce the spin 3/2 projection operator [obtained by using (\ref{eq:114}) to write each spin 3/2 state in terms of a direct product of spin one and spin 1/2 states]
\bea
{\cal P}_{3/2}&=&\sum_{\mu} \left|\sfrac32\,\mu\right> \left<\sfrac32\,\mu\right|
\nonumber\\
&=&\sum_{\mu\nu \nu'} \left<1\,\Lambda_2\,\sfrac12\,\nu | \sfrac32\,\mu\right> \left<1\,\Lambda_1\,\sfrac12\,\nu' | \sfrac32\,\mu\right>
\nonumber\\
&& \qquad\times\left|1\,\Lambda_2\right> \left<1\,\Lambda_1\right|\otimes \left| \sfrac12\,\nu\right> \left< \sfrac12\,\nu'\right|.\qquad
\eea

Now, starting with (\ref{eq:119}), insert this operator between the $\varepsilon^*$ and $D$ (we will show later that the contribution from the spin 1/2 projection operator is zero, and hence inserting the spin 3/2 projection operator is equivalent to inserting unity), use the orthogonality of of the spin one polarization vectors 
($\left|1\,\Lambda\right>={\bm \varepsilon}_{\Lambda}$), 
and insert a complete set of polarization vectors between the $D$ and $\sigma$.
This gives
\begin{align}
&\Theta^{\Lambda\lambda}_{D}({\bf k}_i)= 
\sqrt{\sfrac32}\sum_{\mu\nu'\Lambda'} \left<1\,\Lambda\,\sfrac12\,\nu | \sfrac32\,\mu\right> \left<1\,\Lambda_1\,\sfrac12\,\nu' | \sfrac32\,\mu\right>\left|\sfrac12\,\nu\right>
\nonumber\\
&\qquad\times 
\Big\{({\bm \varepsilon}^*_{\Lambda_1})_\ell
 D^{\ell \ell'}({\bf k}_i)({\bm \varepsilon}_{\Lambda'})_{\ell'}\Big\} 
\left<\sfrac12\,\nu'\right|\,
{\bm \varepsilon}_{\Lambda'}^*   
\cdot\sigma \left|\sfrac12\lambda\right>.
\end{align}  
Use Eq.~(2.20) from Ref.\cite{NDeltaD} 
(derived relativistically, but also true nonrelativistically, and with a sign correction) 
to evaluate the term in curly brackets 
\begin{align}
D_{\Lambda_1,\Lambda'}&\equiv 
({\bm \varepsilon}^*_{\Lambda_1})_{\ell} 
D^{\ell \ell'}( {\bf k}_i)
({\bm \varepsilon}_{\Lambda'})_{\ell} 
 \nonumber\\
 &=- \frac{\sqrt{ 8 \pi}}{3} \; {\bf k}_i^2 
Y_ {2m_\ell} (\hat{{\bf k}}_i)\left<2\, m_\ell\,1\,
\Lambda_1\,|\,1\,\Lambda'\,\right>,   
\end{align}   
(which fixes $\Lambda'=m_\ell+\Lambda_1=m_\ell+\mu-\nu'$, allowing the replacement of the sum over $\Lambda'$ by a sum over $m_\ell$) and use a new identity 
\bea
\left<\sfrac12\,\nu'\right|
{\bm \varepsilon}^*_{\Lambda'}  
\cdot\sigma \left|\sfrac12\lambda\right>=-\sqrt{3}\left<1\Lambda'\sfrac12 \,\nu' |\sfrac12 \lambda\right>  ,
\eea    
to fix $\nu'=\lambda-\Lambda'=\lambda-m_\ell-\mu+\nu'$, which requires that $\mu=\lambda-m_\ell$ and removes the sum over $\mu$.  With these substitutions 
\begin{align}
\Theta^{\Lambda\lambda}_{D}({\bf k}_i)= 
\sqrt{4\pi}\sum_{m_\ell}&{\cal S}_{m_\ell \lambda}\,
{\bf k}_i^2Y_ {2m_\ell} (\hat{{\bf k}}_i)
\nonumber\\
&\times\left<1\,\Lambda\,\sfrac12\,\nu | \sfrac32\,\mu\right>
\left|\sfrac12 \nu\right>  
\end{align}   
where
\begin{align}
{\cal S}_{m_\ell \lambda}=&\sum_{\nu'}\left<2\, m_\ell\,1\,\Lambda_1\,|\,1\,\Lambda'\,\right>\left<1\Lambda'\sfrac12 \nu' |\sfrac12 \lambda\right>
\nonumber\\
&\qquad\times\left<1\,\Lambda_1\,\sfrac12\,\nu' | \sfrac32\,\mu\right>.
\end{align}
This sum can be done using Racah coefficients\footnote{
Rose, Eq.~(6.5a), gives
$$
\sum_{m_2\,{\rm or}\,m_3} \left<j_1 \,m_1\;j_2\,m_2 |j'\right> 
\left<j' \,m_1+m_2\;j_3\,m_3 |j\right>  \left<j_2 \,m_2\;j_3\,m_3 |j''\right>$$
$$= R_{j''j'} \left<j_1 \,m_1\;j''\,m_2+m_3 |j\right>$$
Here we identify
$j_1=2, m_1=m_\ell$,
$j_2=1, m_2=\Lambda_1$,
$j'=1, m_1+m_2=\Lambda'$,
$j_3=\sfrac12, m_3=\nu'$,
$j=\sfrac12$, $m_1+m_2+m_3=\lambda$,
$j''=\sfrac32$, $m_2+m_3=\mu$.}
(from Rose \cite{newreference})
\begin{align}
\sum_{\nu'}\left<2 \,m_\ell\;1\,\Lambda_1 |1\,\Lambda' \right>&
\left<1\,\Lambda'\;\sfrac12 \,\nu' |\sfrac12\,\lambda\right> \left<1\,\Lambda_1 \;\sfrac12 \,\nu'|\sfrac32\,\mu \right> \nonumber\\
&=R_{\frac32\,1}\left<2 \,m_\ell\;\sfrac32\;\mu |\sfrac12\;\lambda\right>. \label{eq:127}
\end{align}
The Racah coefficient, from Table 1.4 on page 227, is
\bea
R_{\frac32\,1}&=&\sqrt{12}\;W(2,1,\sfrac12,\sfrac12; 1, \sfrac32)\nonumber\\
&=&\sqrt{12}\,\sqrt{\frac{5 \,3\, 1\, 1}{4\,3\, 2\,5\,\sfrac12\,\sfrac32\,2}}=1
\eea
Hence, the final answer for $\Theta$ is
\begin{align}
\Theta^{\Lambda \lambda}_D
(\hat {\bf k}_i)=&\sqrt{4\pi}\sum_{m_\ell=-1}^2
\left<2\,m_\ell\,\sfrac32\, \mu | \sfrac12\,\lambda\right> 
{\bf k}_i^2Y_{2m_\ell}(\hat {\bf k}_i) 
\nonumber\\
&\qquad\qquad\times
\left<1\,\Lambda\,\sfrac12\,\nu | \sfrac32\,\mu\right> 
\left| \sfrac12\,\nu\right>,
\end{align}  
which is precisely Eq.~(\ref{eq2.13a}).  

Repeating the derivation using a spin 1/2 projection operator in place of a spin 3/2 operator amounts to replacing the state $\left|\frac32 \mu\right>$ by $\left|\frac12 \mu\right>$ in the sum (\ref{eq:127}), and in this case the Racah coefficient [proportional to $W(2,1,\frac12,\frac12;1,\frac12)$] is zero, proving, as asserted above, that the spin 1/2 contribution is zero.

\section{Nonrelativistic averages over the D-state matrix elements}
\label{apNorma}


Summing over the diquark polarization and averaging over the direction of the relative momentum of the diquark, $r$, using the identities (where $f$ is an arbitrary function)
\bea
\int d^3r\, {\bf r}_i{\bf r}_j \,f({\bf r}^2)&&=
\delta_{ij}\,\frac13 \int d^3r \,{\bf r}^2 f({\bf r}^2)
\nonumber\\
\sum_\Lambda ({\bm \varepsilon}_\Lambda)_{\ell'} 
({\bm \varepsilon}^*_\Lambda)_{\ell}&&=\delta_{\ell'\ell} \label{eq:C1}
\eea   
gives the average of the isospin-0 component 
of the wave function (\ref{eqPsi0a})
\begin{align}
&\left<\left|\Psi^{Da}\right|^2\right>_{\lambda',\lambda}
\equiv \sum_{\Lambda}  \int_{k,r}^{\rm NR}\overline{\Psi}^{Da}_{\Lambda\lambda'}({\bf 0};{\bf k}_1,{\bf k}_2)\Psi^{Da}_{\Lambda\lambda}({\bf 0};{\bf k}_1,{\bf k}_2)
\nonumber\\
& 
=\sfrac34  \int_{k,r}^{\rm NR}\left<\sfrac12\,\lambda'\right|
\sigma_\ell
\sum_{m'}
G^{\ell m'}({\bf k}, {\bf r}) G^{m' m}({\bf k}, {\bf r}) 
\sigma_m  
\left|\sfrac12\,\lambda\right> \Phi^2_D
\nonumber\\
&
=\sfrac53\,\delta_{\lambda'\lambda}  \int_{k,r}^{\rm NR}\,
{\bf k}^2 {\bf r}^2 \; \Phi^2_D(\chi)
\equiv \,\delta_{\lambda'\lambda} \, {\cal N}_D 
 \label{eq:134}
\end{align}   
where the nonrelativistic volume integrals were defined in 
Eq.~(\ref{eq:2.12}), and we assumed that $\Phi_D$ has  the same argument
$\chi= \sfrac12 {\bf k}^2+2{\bf r}^2$
(see Eq.~(\ref{nrarg})) as $\Phi_S$.   
The last line defines the D-state normalization constant.


To evaluate the isospin-1 average, first use the identity
\ba
\sum_\ell 
\int d^3k_i\, 
D^{\ell m'}({\bf k}_i)&&
D^{\ell m}({\bf k}_i) f({\bf k}_i^2)
\nonumber\\
&&=
\delta_{m'm}\,\frac{2}{9}  
\int d^3 k_i \,{\bf k}_i^4 f({\bf k}_i^2),
\ea   
to derive the following convenient result for the normalization 
of the $\Theta$'s
\bea
&&\sum_\Lambda
\int d^3k_i  \, \Theta_{\Lambda \lambda'}^D({\bf k}_i)  \Theta_{\Lambda \lambda}^D
({\bf k}_i) f({\bf k}_i^2)
\nonumber\\
&&=\frac{3}{2}\int d^3k_i \sum_\ell\left<\sfrac12\,\lambda'\right| \sigma_{m'} 
D^{\ell m'}({\bf k}_i)D^{\ell m}({\bf k}_i)\sigma_m \left|\sfrac12\,\lambda\right> f({\bf k}_i^2)
\nonumber\\
&&=\delta_{\lambda'\lambda}\int d^3\,k_i\, {\bf k}_i^4\,f({\bf k}_i^2)\, . \label{eq:C4}
\eea  
Then, summing over the spin-1 diquark polarization vector and using the orthogonality and normalization properties of the $\Theta$'s gives
\begin{align}
\left<\left|\Psi^{Ds}\right|^2\right>_{\lambda',\lambda}
&\equiv\sum_{\Lambda} \int_{k,r}^{\rm NR} \overline{\Psi}^{Ds}_{\Lambda\lambda'}({\bf 0};{\bf k}_1,{\bf k}_2) 
\Psi^{Ds}_{\Lambda \lambda}  
({\bf 0};{\bf k}_1,{\bf k}_2)
\nonumber\\
=
\delta_{\lambda^\prime \lambda} 
& \int_{k,r}^{\rm NR}\,\left(\cos^2\!\phi\; {\bf k}^4 + 
\sin^2\!\phi\; {\bf r}^4\right)\Phi^2_D\, . \label{eqNormaPsi1}
\end{align}   

Now, using the ansatz (\ref{nrarg}) 
(with $\chi=\frac12{\bf k}^2+2{\bf r}^2$ for 
${\bf P}=0$), we can find a relation between the integrals of 
${\bf k}^4, {\bf r}^4$ and ${\bf k}^2{\bf r}^2$ over $\Phi_D$.   
Integrating over ${\bf r}$ by parts first and then over ${\bf k}$, 
gives
\bea
\frac35{\cal N}_D&&=
\left<{\bf k}^2{\bf r}^2\right>\equiv\int_0^\infty 
{\bf k}^2 dk\int_0^\infty {\bf r}^2 dr\, 
({\bf k}^2{\bf r}^2)\, \Phi^2_D(\chi) 
\nonumber\\
&&
=-\sfrac45 \int_0^\infty {\bf k}^4 dk\int_0^\infty dr\, 
{\bf r}^6 \, \frac{d}{d\chi}\Phi^2_D(\chi)  
\nonumber\\
&&=\sfrac{12}5 \int_0^\infty {\bf k}^2 dk\int_0^\infty 
{\bf r}^2 dr\,  ({\bf r}^4)\,\Phi^2_D(\chi)=
\frac{12}5\left<{\bf r}^4\right>.\qquad
\eea  
Doing the same steps, but in reverse order, gives a relation between 
$\left<{\bf k}^4\right>$ and 
$\left<{\bf k}^2 {\bf r}^2\right>$. 
These results can be summarized
\bea
\left< {\bf k}^4\right>&&=
A\left< {\bf k}^2 {\bf r}^2\right>=
\sfrac{20}3\left< {\bf k}^2{\bf r}^2\right>=4\,{\cal N}_D
\nonumber\\
\left< {\bf r}^4\right>&&=B\left< {\bf k}^2 {\bf r}^2\right>=
\sfrac5{12} \left< {\bf k}^2 {\bf r}^2\right>=\sfrac14\,{\cal N}_D \, 
.\label{eq:C7}
\eea  
Hence, both isospin averages can be written in terms of the single normalization constant, ${\cal N}_D$
\bea
\left<\left|\Psi^{Da}\right|^2\right>_{\lambda',\lambda}&&=\delta_{\lambda'\lambda}\,{\cal N}_D
\nonumber\\
\left<\left|\Psi^{Ds}\right|^2\right>_{\lambda',\lambda} 
&&=\delta_{\lambda'\lambda}\,{\cal N}_D \left(4\cos^2\!\phi + \sfrac14\sin^2\!\phi\right).   
\eea   
We choose $\phi$ so that both normalizations are equal, which gives
\bea
\cos\phi=\sqrt{\frac15},\qquad \sin\phi=\sqrt{\frac45}\, .
\eea

Before leaving this section, we point out that the relations (\ref{eq:C7})  
are essential for consistency.  
Using the original definition (\ref{eq:127a}) for $\Psi^{Da}$,  
an alternative calculation 
of the norm using the orthogonality relations (\ref{eq:C4}), gives
\begin{align}
\left<\left|\Psi^{Da}\right|^2\right>_{\lambda',\lambda}
&=\delta_{\lambda'\lambda}\,\sfrac12\Big[\left<{\bf k}_1^4\right> + 
\left<{\bf k}_2^4\right>  \Big]
\nonumber\\
&=\delta_{\lambda'\lambda}\,\sfrac12\Big[\left
<\left(\sfrac12{\bf k}+{\bf r}\right)^4\right>+\left<\left(\sfrac12{\bf k}-{\bf r}\right)^4\right>\Big]
\nonumber\\
&=\delta_{\lambda'\lambda}\,\left<\sfrac1{16} {\bf k}^4+ 
{\bf r}^4+\sfrac56 {\bf k}^2 {\bf r}^2\right>
\nonumber\\
&=\delta_{\lambda'\lambda}\,\left(\sfrac14 +\sfrac14+\sfrac12\right)\,{\cal N}_D
\end{align}  
in agreement with (\ref{eq:134}).

\end{document}